\documentclass[12pt]{article}
\usepackage[utf8]{inputenc}
\usepackage{amsmath}
\usepackage{amssymb}
\usepackage{amsthm}
\usepackage{amsfonts}
\usepackage{graphicx}
\usepackage{multirow}
\usepackage{enumerate}
\usepackage{url} 
\usepackage{float}
\usepackage[table]{xcolor}
\usepackage{colortbl}
\usepackage{tikz}
\usepackage{longtable}
\usepackage{booktabs}
\usepackage{adjustbox}
\usepackage{tabularx}
\usepackage{graphicx}
\usepackage{subcaption} 
\usepackage{authblk}
\usepackage{booktabs}
\usepackage{bigstrut}

\definecolor{rowgray}{RGB}{245,245,235}
\definecolor{rowdark}{RGB}{225,225,215}

\usetikzlibrary{calc}

\usepackage[top=1in,bottom=1.5in,left=1in,right=1in]{geometry}

\usepackage[numbers]{natbib}
\usepackage[unicode=true,bookmarks=true,bookmarksnumbered=false,bookmarksopen=false,breaklinks=false,pdfborder={0 0 1},backref=false,colorlinks=true]{hyperref}
\hypersetup{citecolor=blue}

\title{\Large How should covariates be handled in randomized trials? Empirical evidence from 50 trials and recommendations for practice}
\author{\small{Yulin Shao$^1$, Liangbo Lyu$^1$, Menggang Yu$^1$, Bingkai Wang$^{1,*}$
\vspace{10pt}

$^1$Department of Biostatistics, School of Public Health, University of Michigan, Ann Arbor, MI, USA

$^*$Corresponding author. Address: 1415 Washington Heights, Ann Arbor, MI, USA. Email: bingkai.w@gmail.com}}

\begin{document}
\def\spacingset#1{\renewcommand{\baselinestretch}%
{#1}\small\normalsize} \spacingset{1}

\date{\vspace{-5ex}}

\maketitle
\begin{abstract}
\noindent\textbf{Background and Objective:}
Covariate adjustment can improve precision and power in randomized clinical trials and is recommended by major regulatory agencies. However, there is limited empirical evidence on how different adjustment strategies perform across diverse real-world trials, leaving uncertainty about which methods and covariates should be prespecified in statistical analysis plans. We aim to address this gap and provide practical recommendations. 

\noindent\textbf{Methods:}
We conducted a large-scale empirical study using individual-level data from 50 publicly available randomized trials (29,094 participants; 574 treatment-outcome comparisons). We compared commonly used covariate-adjusted estimators, including analysis of covariance, inverse-probability weighting, g-computation, and machine-learning-based approaches, combined with three covariate-selection strategies. Performance was evaluated using precision gains, changes in point estimates, computational reliability, and the probability that covariate adjustment altered statistical significance relative to an unadjusted analysis.

\noindent\textbf{Results:}
Covariate adjustment improved precision in most settings, with a median variance reduction of 13.3\% for continuous outcomes and 4.6\% for binary outcomes. Adjustment was substantially more likely to produce a gain in statistical significance than a loss. Parsimonious regression approaches using a small prespecified set of prognostic covariates performed as well as or better than more complex methods, particularly in small to medium samples. Machine-learning-based estimators did not provide additional precision and were more prone to computational failure for binary outcomes.

\noindent\textbf{Conclusions:}
Across a wide range of randomized trials, parsimonious covariate adjustment provided consistent efficiency gains without introducing systematic bias. These findings support routine covariate adjustment in primary trial analyses and provide practical guidance for writing statistical analysis plans, selecting covariates, and planning sample size. All curated datasets and analysis code are openly released as a reproducible resource to support future clinical research.
\end{abstract}
\noindent%


\spacingset{1.5}

\section{Introduction}
In randomized clinical trials (RCTs), incorporating baseline covariates into the primary analysis (commonly referred to as covariate adjustment) has long been recognized as a principled way to improve precision and power without compromising the validity of treatment effect estimation. This view is supported by methodological research \cite{kahan2012reporting, ciolino2019ideal, morris2022planning, van2024covariate}, regulatory guidance from the U.S. Food and Drug Administration \cite{FDA2023covariate} and the European Medicines Agency \cite{EMA2015baseline}, and recent recommendations on estimands and statistical analysis plans \citep{ICH2020E9R1, kahan2024estimands}. Despite this broad consensus, several practical questions remain unresolved: Which covariates should be included? How complex should the adjustment model be? And to what extent do different adjustment strategies change the conclusion of trials?

These questions arise routinely during trial design and the development of statistical analysis plans. The choice of adjustment model and covariates has direct implications for efficiency, robustness, and reproducibility.
More flexible approaches, such as regression models with interaction terms \cite{Tsiatis2008} or machine-learning-based estimators \cite{van2011targeted}, can capture nonlinear relationships and treatment-effect heterogeneity and may therefore improve precision.
However, simpler models with a small prespecified set of covariates can offer greater stability and avoid overfitting in finite samples \citep{subramanian2013overfitting}.
Clear empirical evidence comparing these approaches in real trials is therefore essential for informing best practice.

Most existing evidence on covariate adjustment is derived from asymptotic theory, simulation studies, or individual case studies. Although these approaches have generated important insights, they do not fully reflect the heterogeneity of sample sizes, covariate structures, and outcome-covariate relationships encountered in practice. Simulation results \cite{kahan2014risks, chausse2016simulation, gao2024does} depend on assumed data-generating mechanisms, and isolated case studies \cite{thompson2015covariate, kahan2016comparison} are limited in scope and generalizability. As a result, trialists and applied statisticians still lack practice-oriented guidance on how covariate adjustment performs across diverse real-world settings and how it should be prespecified for primary analyses. To address this gap, we conducted a large-scale empirical evaluation of commonly used covariate-adjustment strategies and provided practical recommendations for their use in RCTs.

\section{Methods}
\subsection{Data curation}
We searched four public data repositories, Dryad (\url{https://datadryad.org}), Harvard Dataverse (\url{https://dataverse.harvard.edu}), Zenodo (\url{https://zenodo.org}), and the R CRAN repository (\url{https://cran.r-project.org}) with keywords ``randomized clinical trials'', ``randomized trials'', and ``randomized controlled trials''. Our objective was to identify 50 RCTs with accessible individual-level data, well-defined data structures, and accompanying data dictionaries. We excluded studies that did not focus on human health outcomes, were not individually randomized, had unclear data dictionary, lacked an associated peer-reviewed publication, or were completed before 1990. This search was designed to assemble a broad and reusable benchmark dataset rather than to provide an exhaustive systematic review of all RCTs. 
As a retrospective empirical study based on publicly available trial datasets, this work did not have a prespecified protocol, and no finalized reporting guideline was identified for studies of this type.

For eligible RCTs, we implemented a unified data-processing pipeline. Primary and secondary outcomes were defined according to the trial registry when available, or otherwise based on the primary publication. Treatment assignment variables were harmonized across studies. 
{For each trial, we constructed a candidate pool of baseline covariates using variables prespecified in the primary statistical analysis, when available, and variables reported in the baseline characteristics table of the primary publication.  The covariate-adjustment strategies were then applied to this common candidate pool to enable consistent comparisons of adjustment methods across trials.}
Outcome variables with more than 40\% missingness were excluded. Observations with missing values in the remaining outcome variables were omitted from the analysis. Missing baseline covariates were imputed using the mean for continuous variables and the mode for categorical variables. 

\subsection{Statistical methods}
{We compared covariate-adjusted estimators of the marginal average treatment effect for continuous outcomes and the marginal risk difference for binary outcomes.} The estimators were grouped into two broad families.

The first family comprised commonly used regression-based approaches: analysis of covariance (ANCOVA) \citep{YangTsiatis2001}, analysis of heterogeneous covariance (ANHECOVA) \citep{Tsiatis2008,ye2023toward}, inverse-probability weighting (IPW) \citep{williamson2014variance}, and g-computation using logistic regression for binary outcomes (g-logistic) \citep{Moore2009a}. These methods are widely applied in primary analyses of clinical trials and are compatible with standard statistical analysis plans.

The second family consisted of estimators that incorporate machine-learning-based outcome models. We included these approaches because they are increasingly advocated in the methodological literature for improving efficiency in RCTs.  We considered targeted minimum loss estimation (TMLE) \citep{van2011targeted} and debiased machine learning (DML) \citep{chernozhukov2018double}. For outcome prediction, both approaches used an ensemble of generalized linear models, penalized regression (GLMNET) \cite{hastie2015statistical}, random forests \cite{breiman2001random}, and Bayesian additive regression trees \cite{chipman2010bart}, with candidate algorithms selected based on predictive performance.

All covariate-adjusted estimators were compared with the unadjusted estimator, defined as the difference in mean outcomes between randomized groups. Technical details, software implementation, and algorithm selection are provided in Supplementary Material B.

For each covariate-adjusted method, we examined three covariate-selection strategies: \textit{All}, \textit{Top-3}, and \textit{Baseline+}. The \textit{All} strategy adjusted for all available baseline variables, maximizing the information used for prognostic adjustment but risking numerical instability when the covariate dimension is large relative to sample size. The \textit{Top-3} strategy adjusted only for the three covariates most strongly correlated with the outcome, representing a data-adaptive approach that prioritizes prognostic strength while reducing model complexity. The \textit{Baseline+} strategy used a prespecified set of commonly recommended variables, including the baseline outcome, stratification factors, age, sex, and weight when available, providing a transparent adjustment set. Table~\ref{tab:methods} summarizes all considered estimators and covariate-selection strategies.

\begin{table}[p]
\caption{Statistical models and covariate selection strategies evaluated in this study. }
\label{tab:methods}
\centering

\rowcolors{2}{white}{rowgray}
\resizebox{\textwidth}{!}{
\begin{tabular}{p{3.2cm} p{13.5cm}}

\rowcolor{rowdark}
\hline
\multicolumn{2}{l}{\textbf{Estimators}} \\
\hline

Unadjusted & Difference in mean outcomes between groups. No covariates involved.\\

ANCOVA  & Classical regression adjusting for baseline covariates; estimates the treatment effect from a linear model including treatment and covariates. \\

ANHECOVA & Extension of ANCOVA that allows treatment--covariate interactions; implemented using g-computation for improved flexibility. \\

IPW  & Outcome comparison after reweighting individuals by the estimated probability of receiving their observed treatment based on a logistic model. \\

g-logistic  & G-computation estimator obtained from a logistic regression of the outcome on treatment and covariates. \\

DML  &  Cross-fitted outcome and treatment models are combined using an influence-function-based estimator. \\

TMLE  & Machine-learning-based predictions are refined via a targeted update step to yield a g-computation estimator. \\

\rowcolor{rowdark}
\hline
\multicolumn{2}{l}{\textbf{Covariate-selection strategies}} \\
\hline

All & Adjust for all available baseline covariates. \\

Top-3 & {Adjust for the three baseline covariates with the largest absolute Pearson correlations with the outcome, computed within each trial after pooling treatment groups.}  \\

Baseline+ & Adjust for a prespecified set of commonly recommended variables, including the baseline outcome, stratification factors, age, sex, and weight. \\
\hline
\end{tabular}
}
\end{table}

\subsection{Performance metrics}
We evaluated performance using measures that reflect the practical consequences of covariate adjustment for trial analysis. All metrics were computed for each treatment-outcome comparison.

\textbf{Precision.} Precision gain was quantified using the proportional variance reduction (PVR) relative to the unadjusted estimator. This measure represents the percentage reduction in the variance of the treatment effect estimates and can be interpreted as the corresponding reduction in the required sample size to achieve the same statistical power. Positive values therefore indicate increased efficiency from covariate adjustment.


\textbf{Change in point estimates.} To assess whether covariate adjustment systematically altered the estimated treatment effect, we calculated the difference between adjusted and unadjusted point estimates, scaled by the standard error of the unadjusted estimator. Values centered around zero indicate that adjustment introduces no systematic shifts in the estimand.


\textbf{Operating characteristics for trial conclusions.}
To summarize the real-world impact of covariate adjustment on statistical inference, we evaluated three quantities: the error rate, the covariate adjustment gain (CAG), and the covariate adjustment loss (CAL). The error rate was defined as the proportion of analyses in which the software failed to return a result.
CAG is the probability that a covariate-adjusted analysis yields a statistically significant treatment effect {(two-sided tests of zero effect at the 0.05 level)} when the corresponding unadjusted analysis does not, whereas CAL is the probability that statistical significance is lost after adjustment. These quantities directly describe how often covariate adjustment would change the statistical conclusion of a trial. 


\section{Results}
\subsection{Data curation results}
We identified 50 eligible RCTs, including 38 from Dryad, 2 from Harvard Dataverse, 2 from Zenodo, and 8 from the R CRAN repository. These trials comprised 29,094 participants and 574 treatment-outcome comparisons, of which 445 involved continuous outcomes and 129 involved binary outcomes.

\begin{figure}
    \centering
    \includegraphics[width=1\linewidth]{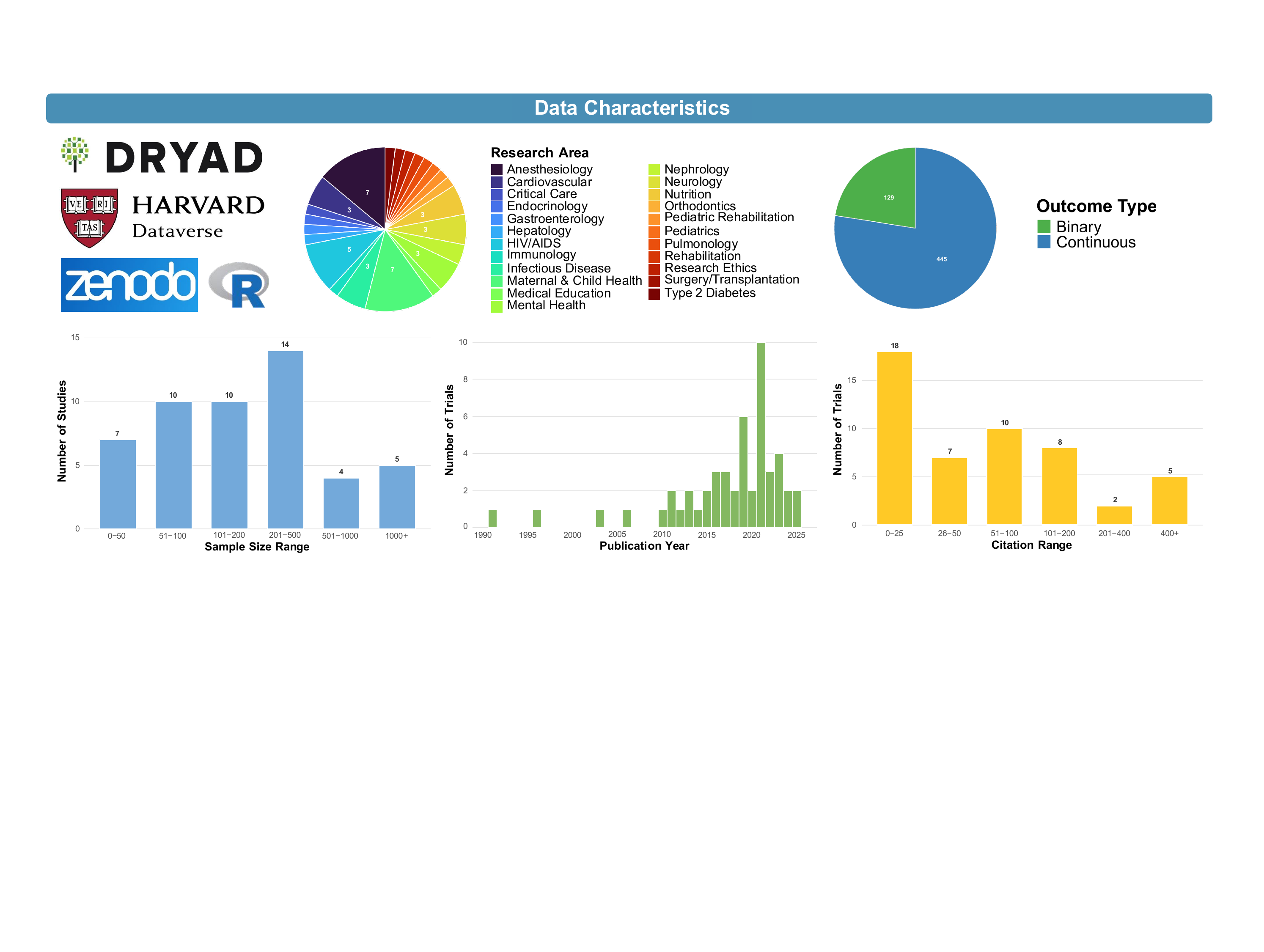}
    \vspace{-6cm}
    \caption{Characteristics of curated RCTs, {including data sources, research areas, outcome types, sample sizes, publication years, and citation counts.}}
    \label{fig:data-char}
\end{figure}

Figure~\ref{fig:data-char} summarizes the characteristics of the curated dataset. The included trials span a wide range of therapeutic areas and sample sizes, were predominantly conducted within the past 15 years, and have been widely cited in the literature, indicating that the dataset reflects contemporary and clinically relevant trial settings. Trial-level details are provided in Supplementary Table 1.

\subsection{Precision}
\begin{figure}
    \centering
    \includegraphics[width=1\linewidth]{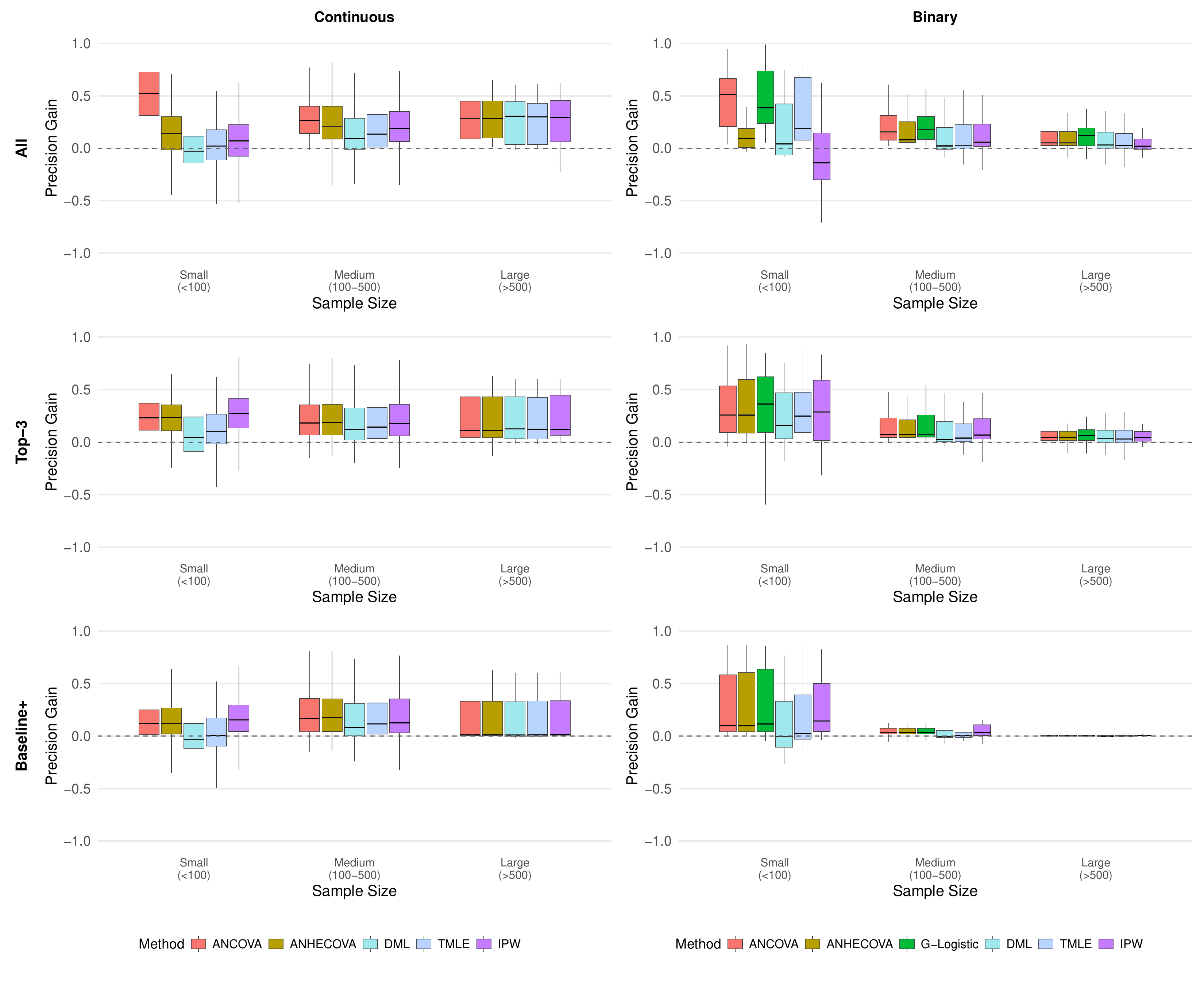}
    \caption{Box plots showing {the empirical distribution of} percentage precision gains across different sample-size categories. In the six panels, columns separate outcome types and rows represent covariate-selection strategies, while colors denote the adjustment methods. The y-axis shows the proportional variance reduction (PVR) relative to the unadjusted analysis, with positive values indicating improved precision. The x-axis categorizes trials by sample size: small (0-100), medium (100-500), and large ($>$500). {Other abbreviations used in the figure are defined in Table~\ref{tab:methods}.}}
    \label{fig:precision}
\end{figure}

Figure \ref{fig:precision} summarizes the proportional variance reduction achieved by each covariate-adjustment strategy. Across all methods and sample sizes, covariate adjustment improved precision on average, with a median variance reduction of 13.3\% for continuous outcomes and 4.6\% for binary outcomes. These improvements translate directly into proportional reductions in the sample size required to achieve a desired statistical power.


For continuous outcomes adjusting for all covariates, ANCOVA yielded the largest precision gains and rarely produced losses. These benefits were especially pronounced in small samples, where complex models were vulnerable to overfitting. ANHECOVA and IPW performed similarly to ANCOVA in medium to large samples but occasionally lost precision in smaller trials, likely because the inclusion of treatment-covariate interaction terms reduces effective degrees of freedom and increases variance. 
Machine-learning-based estimators (TMLE and DML) approached the performance of ANCOVA and ANHECOVA in large samples but were generally less precise in small samples.

When adjustment was restricted to the three most prognostic covariates (Top-3 strategy), the instability observed under the all-covariates strategy was reduced, and the performance of ANCOVA, ANHECOVA, and IPW became similar across sample sizes. In contrast, the relative performance of TMLE and DML changed little under this strategy, indicating that excluding weakly prognostic covariates had limited influence on the efficiency of these estimators.

The Baseline+ strategy produced smaller precision gains than adjusting for all covariates and slightly smaller variance reductions than Top-3, consistent with its more conservative prespecified design. Nevertheless, overall precision improvements remained consistently positive across sample sizes, and differences between estimators became minimal in medium to large trials.

For binary outcomes, the overall patterns were similar, but the magnitude of precision gains was smaller. This likely reflects the more limited information content of binary outcomes compared with continuous measures. Despite this attenuation, ANCOVA combined with the Baseline+ covariate set still achieved a median variance reduction of 10.6\%, demonstrating {practically relevant efficiency gains} even for binary endpoints. In addition, g-logistic regression performed comparably to ANCOVA across the curated datasets, suggesting that both approaches can provide substantial precision improvements in practice.

Supplementary Figure 2 presents parallel analyses restricted to the primary outcomes of each RCT and shows similar results.

\subsection{Point estimates}
\begin{figure}
    \centering
    \includegraphics[width=1\linewidth]{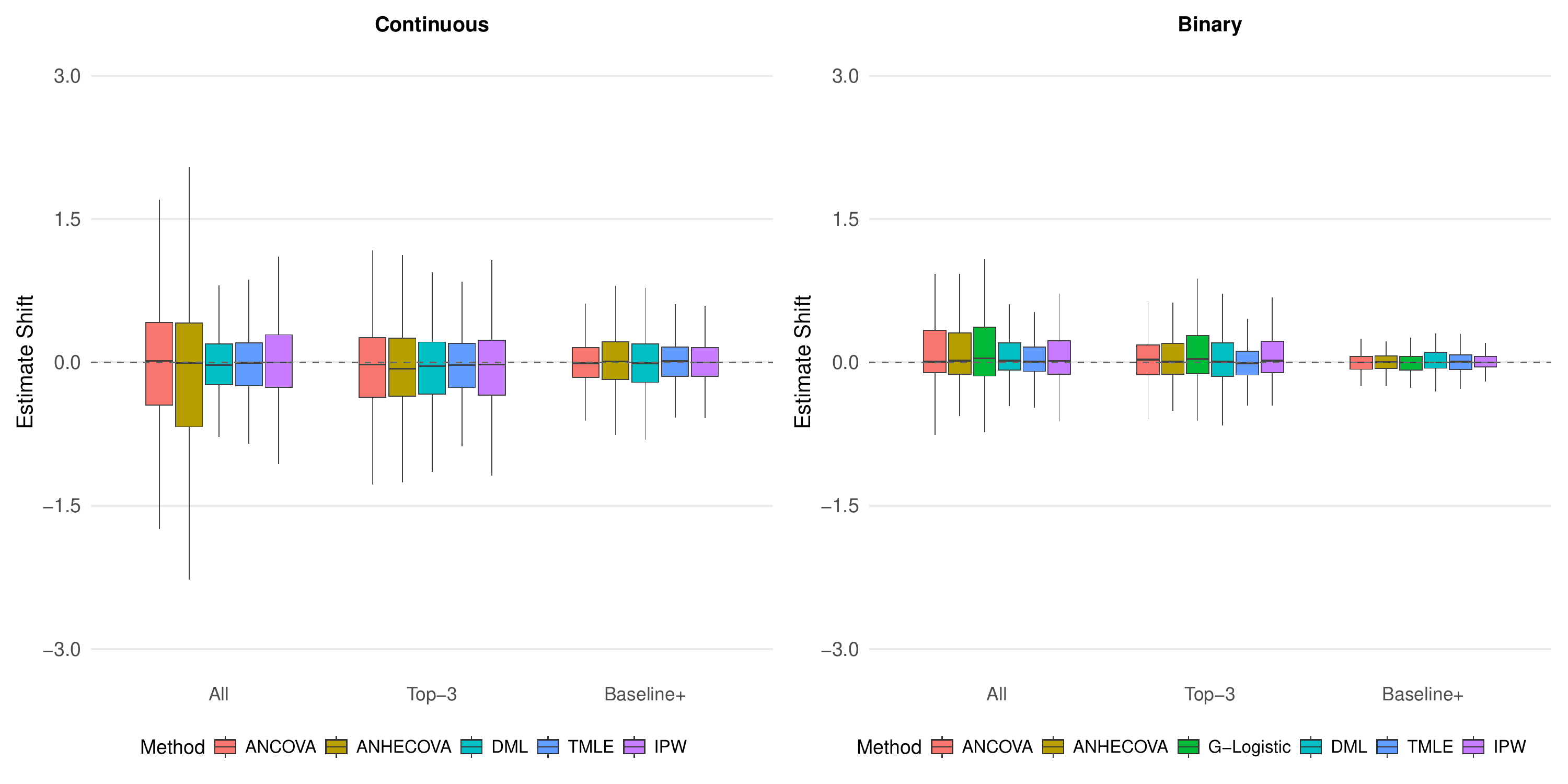}
    \caption{Box plots showing {the empirical distribution of} estimate shifts scaled by the standard error of the unadjusted estimators across adjustment methods and covariate-selection strategies. {Column panels separate outcome types, and abbreviations used in the figure are defined in Table~\ref{tab:methods}.}}
    \label{fig:est-shift}
\end{figure}

Point-estimate comparisons are summarized in Figure~\ref{fig:est-shift}. Across all settings, the point estimate shifts were approximately symmetric around zero, suggesting no systematic bias was introduced by the covariate adjustment methods. Compared with adjusting for all covariates, more conservative approaches (Top-3 and Baseline+) yielded more stability in estimate shifts. 
In addition, the adjusted and unadjusted estimators exhibited closer point estimates for binary outcomes than for continuous outcomes. {This pattern may reflect the bounded nature of binary outcomes, which leaves less room for point-estimate shifts, and the weaker outcome variation explained by baseline covariates in our empirical datasets.}

Differences in point estimates generally arise from multiple sources and may not necessarily indicate bias. Taking ANCOVA as an illustrative example, Wang et al. \cite{wang2019analysis} showed that ANCOVA approximately subtracts from the unadjusted estimate the component of the treatment effect driven by covariate imbalance and its prognostic strength. Consequently, when the adjusted covariates are well balanced or weakly prognostic, the discrepancy between the ANCOVA and unadjusted estimators tends to be small. For other estimators, the scaled difference may also depend on additional factors such as treatment-effect heterogeneity and the allocation ratio.

\subsection{Covariate adjustment gains and losses}

As summarized in Table~\ref{tab:cag_cal}, covariate adjustment was generally advantageous across methods and outcome types: CAG consistently exceeded CAL, indicating that adjustment is more likely to create gains than to produce losses. For continuous outcomes, the largest benefits were observed for ANCOVA with all covariates (CAG = 17\%, CAL = 4\%), followed by ANHECOVA and IPW. For binary outcomes, gains were smaller overall, with the largest CAG observed for g-logistic (CAG = 11\%, CAL = 2\%). Machine-learning-based estimators showed modest gains but low loss rates.

\begin{table}[htbp]
\centering
\caption{Covariate adjustment gain (CAG) and loss (CAL) across methods and covariate-selection strategies. {Abbreviations for covariate-adjustment strategies are defined in Table~\ref{tab:methods}.}}
\label{tab:cag_cal}
\begin{tabular}{llrrrrrr}
\hline
\rowcolor{rowdark}
Outcome & Method & \multicolumn{2}{c}{All} & \multicolumn{2}{c}{Top 3} & \multicolumn{2}{c}{Baseline+} \\
\cline{3-8}
\rowcolor{rowdark}
 &   & CAG & CAL & CAG & CAL & CAG & CAL \\
 \hline
 
\multirow{5}{*}{Continuous} & ANCOVA     & 17\% & 4\%  & 6\% & 2\% & 6\% & 2\% \\
                             & ANHECOVA   & 12\% & 7\%  & 7\% & 2\% & 6\% & 2\% \\
                             & IPW        & 6\%  & 6\%  & 7\% & 2\% & 6\% & 2\% \\
                             & DML        & 7\%  & 4\%  & 6\% & 3\% & 6\% & 4\% \\
                             & TMLE       & 5\%  & 3\%  & 5\% & 2\% & 5\% & 2\% \\
\hline
\multirow{6}{*}{Binary}     & ANCOVA     & 8\%  & 1\%  & 3\% & 2\% & 3\% & 1\% \\
                             & ANHECOVA   & 4\%  & 1\%  & 3\% & 2\% & 3\% & 0\% \\
                             & IPW        & 3\%  & 0\%  & 3\% & 0\% & 3\% & 0\% \\
                             & G-logistic & 11\% & 2\%  & 7\% & 2\% & 4\% & 1\% \\
                             & DML        & 4\%  & 0\%  & 3\% & 1\% & 2\% & 0\% \\
                             & TMLE       & 1\%  & 1\%  & 2\% & 1\% & 2\% & 0\% \\
\hline
\end{tabular}
\end{table}

Patterns across covariate-selection strategies reveal a trade-off between maximizing efficiency and controlling instability. Adjusting for all covariates produced the highest CAG values but also slightly elevated CAL, consistent with increased sensitivity to small-sample variability. In contrast, conservative prespecified strategies such as Baseline+ yielded smaller but stable gains: across all methods, covariate adjustment under Baseline+ remained roughly three times more likely to yield a gain than a loss.
Notably, some apparent losses may reflect the removal of spurious statistical significance arising from baseline imbalance, as discussed in Section~3.3. 

\subsection{Computation error rate}
Table~\ref{tab:error_rates} summarizes the computational error rates for each method. {Algorithm-specific non-convergence warnings were partially reflected in the error rates and not separately reported.}
For continuous outcomes, all covariate-adjustment approaches ran reliably and returned valid results for every treatment-outcome comparison. In contrast, several methods exhibited non-negligible error rates for binary outcomes. 

\begin{table}[htbp]
\centering
\caption{Error rates across methods and covariate-selection strategies. {Abbreviations for covariate-adjustment strategies are defined in Table~\ref{tab:methods}.}}
\label{tab:error_rates}
\begin{tabular}{llrrr}
\hline
\rowcolor{rowdark}
Outcome & Method & All & Top 3 & Baseline+ \\
\hline
\multirow{5}{*}{Continuous} & ANCOVA     & 0\% & 0\% & 0\% \\
                             & ANHECOVA   & 0\% & 0\% & 0\% \\
                             & IPW        & 0\% & 0\% & 0\% \\
                             & DML        & 0\% & 0\% & 0\% \\
                             & TMLE       & 0\% & 0\% & 0\% \\
\hline
\multirow{6}{*}{Binary}     & ANCOVA     & 0\%  & 0\%  & 0\% \\
                             & ANHECOVA   & 0\%  & 0\%  & 0\% \\
                             & IPW        & 3\%  & 2\%  & 0\% \\
                             & G-logistic & 4\%  & 1\%  & 0\% \\
                             & DML        & 12\% & 12\% & 8\% \\
                             & TMLE       & 4\%  & 4\%  & 3\% \\
\hline
\end{tabular}
\end{table}

Most computational failures were associated with sparse binary outcomes. When using logistic regression, rare outcome categories combined with a large number of covariates occasionally resulted in near-singular covariance matrices, yielding unstable standard error estimates or convergence failures. For DML and TMLE, since they rely on cross-fitting, rare events may be absent within training folds, leading to model-fitting failure. TMLE was comparatively more stable because its widely used R implementation incorporates safeguards and fallback procedures that mitigate fitting failures. 
Overall, these findings indicate that covariate adjustment for binary outcomes is more susceptible to computational instability than for continuous outcomes, particularly in the presence of rare events.


\section{Discussion}
As covariate adjustment becomes increasingly common in RCT analyses, investigators need practical guidance on choosing adjustment methods and covariate-selection strategies. We addressed this need through a large-scale empirical benchmarking study using individual-level data from 50 completed RCTs. By analyzing observed trial data, our evaluation captures features difficult to represent in simulations and provides practice-oriented evidence to support recommendations for primary trial analyses.

Our findings showed that flexible, data-adaptive algorithms did not outperform classical parsimonious regression models for covariate adjustment in RCTs. A likely explanation is that many RCT datasets lack the structural complexity required for machine-learning methods to deliver efficiency gains. Randomization reduces the need for extensive modeling of confounding, treatment effects are often well approximated by relatively simple outcome models, and typical trial sample sizes limit the stability of highly flexible estimators. Although these considerations are specific to randomized settings, similar patterns have been observed in observational studies, where simple regression approaches frequently match or exceed machine-learning methods in terms of the bias-variance trade-off \citep{keele2021comparing, doutreligne2023select, chen2025we}. Concerns about the finite-sample reliability of data-adaptive estimators have also been noted in both methodological and applied work \citep{mooney2021thirteen, zivich2021machine, hines2022demystifying, wang2023CRT}. 
At the same time, machine-learning methods remain valuable tools within the broader causal inference framework. They are particularly well-suited to settings with complex data structures or when the primary objective is to estimate heterogeneous treatment effects \citep{baiardi2024value}. The key challenge is therefore not to replace simple models with more complex ones, but to identify settings in which their advantages align with the available data.

\subsection{Practical recommendations}
Our analyses provide clear insights for the prespecification and implementation of covariate adjustment in RCTs.

First, we recommend adjustment for prognostic baseline covariates to enhance statistical efficiency and power. 
These gains were consistently larger for continuous outcomes than for binary outcomes. Importantly, the probability that covariate adjustment led to a gain in statistical significance exceeded the probability of a loss, indicating a net benefit in practice.


Second, we recommend parsimonious regression methods, such as ANCOVA, for covariate adjustment in primary analyses. 
In our empirical evaluation, increased model flexibility did not translate into {practically relevant} precision gains. Differences between estimators diminished as sample size increased, while in smaller trials simpler models consistently performed better because of their numerical stability and robustness. 
{For variance estimation, we recommend using robust (sandwich) standard errors that provide asymptotically valid inference even when the working outcome model is misspecified \citep{Tsiatis2006, FDA2023covariate}. However, we note that sandwich variance estimators may be inaccurate in small samples, and their finite-sample performance can depend on the design and correction methods \citep{kauermann2001note}.}

Third, we recommend prespecifying a small set of clinically established prognostic covariates.
The Baseline+ strategy produced precision gains comparable to those of data-driven selection while avoiding the risks of post hoc variable selection and maintaining regulatory transparency. This approach is broadly applicable across therapeutic areas and can be tailored using subject-matter knowledge. For example, in diabetes trials, duration of diabetes and glycated hemoglobin (HbA1c) are well-established prognostic factors and natural candidates for additional adjustment \citep{rawshani2019relative}. Although adjusting for all available covariates yielded the largest gains in some settings, it was more prone to computational instability in small samples.

\begin{figure}
    \centering
    \includegraphics[width=\linewidth]{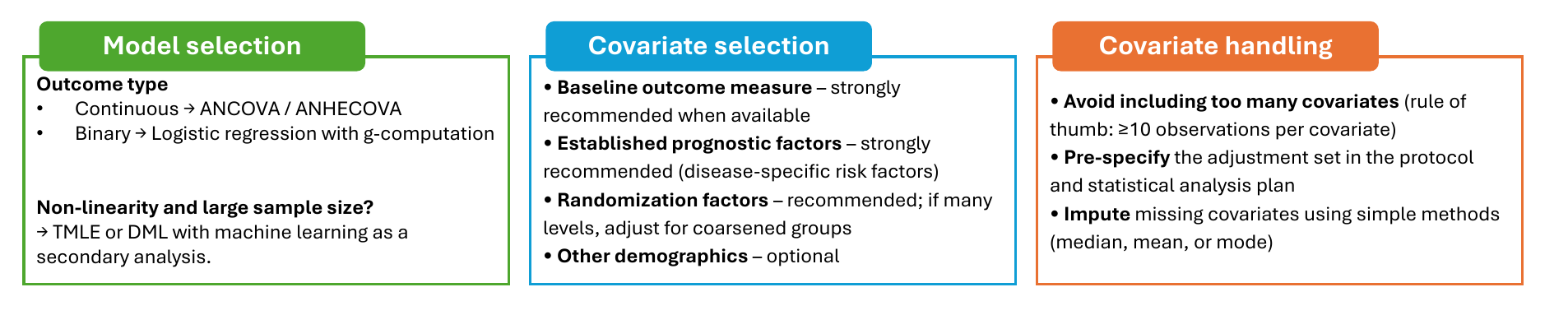}

\vspace{-0.2in}
    
    \caption{Practical recommendations for covariate adjustment.}
    \label{fig:recommendation}
\end{figure}

Figure~\ref{fig:recommendation} summarizes these recommendations as a practical workflow for choosing adjustment models, selecting covariates, and handling covariates in statistical analysis plans. Across settings, the key principles are prespecification, parsimony, and prioritization of established prognostic variables.

\subsection{Limitations}
This study has several limitations. First, the curated dataset was assembled from a non-exhaustive search of publicly available RCTs. Although it spans multiple therapeutic areas and a wide range of sample sizes and designs, it may not fully represent the entire spectrum of modern clinical trials. In addition, many of the 574 treatment-outcome comparisons were derived from the same trials and therefore share baseline covariate information, inducing correlation among observations. However, analyses restricted to primary outcomes (Supplementary Figure 2) yielded consistent results. Accordingly, our findings should be interpreted as evidence of broad empirical patterns rather than predictions for any specific trial or disease setting.

Second, we evaluated a focused set of covariate-adjustment methods to facilitate transparent comparisons. However, other approaches may perform better under particular conditions. For example, overlap weighting has been shown to improve stability relative to traditional inverse-probability weighting \citep{li2018balancing}. {We additionally evaluated these types of methods and presented the results in the Supplementary Material.} Hyperparameter tuning may enhance the performance of machine-learning-based estimators \citep{chernozhukov2018double}. In RCTs, however, such tuning is often constrained by prespecification requirements and the need for reproducible analysis plans.

Third, our analysis focused on marginal average treatment effects for continuous and binary outcomes.  Although this estimand is central to most regulatory and confirmatory trials, alternative estimands, such as conditional average treatment effects, marginal risk ratios, odds ratios, or survival-based measures, may interact with covariate adjustment differently \citep{hernan2020causal}. Extending empirical evaluation to these settings would provide a more comprehensive understanding of the operating characteristics of adjustment strategies.

Finally, our evaluation did not incorporate several common design and data features, including missing outcomes and covariate-adaptive randomization. Missing data are ubiquitous in clinical research and are typically addressed using principled methods combined with sensitivity analyses to assess robustness \citep{little2019statistical, daniels2008missing}. Covariate-adaptive randomization procedures, such as stratified randomization, are also widely used and can improve efficiency when paired with appropriate covariate adjustment \citep{bugni2018inference, wang2023model, bannick2025general}. Incorporating these elements into future empirical studies will further clarify how design features and adjustment strategies interact in practice.

\section*{Data availability}
Source data are provided with this paper. All data used in our research come from publicly available sources. All preprocessed datasets are available at \url{https://github.com/syl051088/RCT_Bench} with unrestricted access. We encourage researchers to use these resources for benchmarking, replication, and further clinical research.


{
\bibliographystyle{unsrtnat}
\bibliography{references}

\begin{thebibliography}{43}
\providecommand{\natexlab}[1]{#1}
\providecommand{\url}[1]{\texttt{#1}}
\expandafter\ifx\csname urlstyle\endcsname\relax
  \providecommand{\doi}[1]{doi: #1}\else
  \providecommand{\doi}{doi: \begingroup \urlstyle{rm}\Url}\fi

\bibitem[Kahan and Morris(2012)]{kahan2012reporting}
Brennan~C Kahan and Tim~P Morris.
\newblock Reporting and analysis of trials using stratified randomisation in leading medical journals: review and reanalysis.
\newblock \emph{Bmj}, 345, 2012.

\bibitem[Ciolino et~al.(2019)Ciolino, Palac, Yang, Vaca, and Belli]{ciolino2019ideal}
Jody~D Ciolino, Hannah~L Palac, Amy Yang, Mireya Vaca, and Hayley~M Belli.
\newblock Ideal vs. real: a systematic review on handling covariates in randomized controlled trials.
\newblock \emph{BMC medical research methodology}, 19\penalty0 (1):\penalty0 136, 2019.

\bibitem[Morris et~al.(2022)Morris, Walker, Williamson, and White]{morris2022planning}
Tim~P Morris, A~Sarah Walker, Elizabeth~J Williamson, and Ian~R White.
\newblock Planning a method for covariate adjustment in individually randomised trials: a practical guide.
\newblock \emph{Trials}, 23\penalty0 (1):\penalty0 328, 2022.

\bibitem[Van~Lancker et~al.(2024)Van~Lancker, Bretz, and Dukes]{van2024covariate}
Kelly Van~Lancker, Frank Bretz, and Oliver Dukes.
\newblock Covariate adjustment in randomized controlled trials: General concepts and practical considerations.
\newblock \emph{Clinical Trials}, 21\penalty0 (4):\penalty0 399--411, 2024.

\bibitem[{FDA}(2023)]{FDA2023covariate}
{FDA}.
\newblock Adjusting for covariates in randomized clinical trials for drugs and biological products: Guidance for industry.
\newblock \url{https://www.fda.gov/media/148910/download}, 2023.

\bibitem[{EMA}(2015)]{EMA2015baseline}
{EMA}.
\newblock Guideline on adjustment for baseline covariates in clinical trials, 2015.

\bibitem[{ICH}(2020)]{ICH2020E9R1}
{ICH}.
\newblock {ICH9 E9 (R1)} addendum on estimands and sensitivity analysis in clinical trials to the guideline on statistical principles for clinical trials, 2020.

\bibitem[Kahan et~al.(2024)Kahan, Hindley, Edwards, Cro, and Morris]{kahan2024estimands}
Brennan~C Kahan, Joanna Hindley, Mark Edwards, Suzie Cro, and Tim~P Morris.
\newblock The estimands framework: a primer on the ich e9 (r1) addendum.
\newblock \emph{bmj}, 384, 2024.

\bibitem[Tsiatis et~al.(2008)Tsiatis, Davidian, Zhang, and Lu]{Tsiatis2008}
A.A. Tsiatis, M.~Davidian, M.~Zhang, and X.~Lu.
\newblock Covariate adjustment for two-sample treatment comparisons in randomized clinical trials: A principled yet flexible approach.
\newblock \emph{Stat Med}, 27\penalty0 (23):\penalty0 4658--4677, 2008.

\bibitem[Van~der Laan et~al.(2011)Van~der Laan, Rose, et~al.]{van2011targeted}
Mark~J Van~der Laan, Sherri Rose, et~al.
\newblock \emph{Targeted learning: causal inference for observational and experimental data}, volume~4.
\newblock Springer, 2011.

\bibitem[Subramanian and Simon(2013)]{subramanian2013overfitting}
Jyothi Subramanian and Richard Simon.
\newblock Overfitting in prediction models--is it a problem only in high dimensions?
\newblock \emph{Contemporary clinical trials}, 36\penalty0 (2):\penalty0 636--641, 2013.

\bibitem[Kahan et~al.(2014)Kahan, Jairath, Dor{\'e}, and Morris]{kahan2014risks}
Brennan~C Kahan, Vipul Jairath, Caroline~J Dor{\'e}, and Tim~P Morris.
\newblock The risks and rewards of covariate adjustment in randomized trials: an assessment of 12 outcomes from 8 studies.
\newblock \emph{Trials}, 15\penalty0 (1):\penalty0 139, 2014.

\bibitem[Chauss{\'e} et~al.(2016)Chauss{\'e}, Liu, and Luta]{chausse2016simulation}
Pierre Chauss{\'e}, Jin Liu, and George Luta.
\newblock A simulation-based comparison of covariate adjustment methods for the analysis of randomized controlled trials.
\newblock \emph{International Journal of Environmental Research and Public Health}, 13\penalty0 (4):\penalty0 414, 2016.

\bibitem[Gao et~al.(2024)Gao, Liu, and Matsouaka]{gao2024does}
Ying Gao, Yi~Liu, and Roland Matsouaka.
\newblock When does adjusting covariate under randomization help? a comparative study on current practices.
\newblock \emph{BMC Medical Research Methodology}, 24\penalty0 (1):\penalty0 250, 2024.

\bibitem[Thompson et~al.(2015)Thompson, Lingsma, Whiteley, Murray, and Steyerberg]{thompson2015covariate}
Douglas~D Thompson, Hester~F Lingsma, William~N Whiteley, Gordon~D Murray, and Ewout~W Steyerberg.
\newblock Covariate adjustment had similar benefits in small and large randomized controlled trials.
\newblock \emph{Journal of clinical epidemiology}, 68\penalty0 (9):\penalty0 1068--1075, 2015.

\bibitem[Kahan et~al.(2016)Kahan, Rushton, Morris, and Daniel]{kahan2016comparison}
Brennan~C Kahan, Helen Rushton, Tim~P Morris, and Rhian~M Daniel.
\newblock A comparison of methods to adjust for continuous covariates in the analysis of randomised trials.
\newblock \emph{BMC medical research methodology}, 16\penalty0 (1):\penalty0 42, 2016.

\bibitem[Yang and Tsiatis(2001)]{YangTsiatis2001}
L.~Yang and A.A. Tsiatis.
\newblock Efficiency study of estimators for a treatment effect in a pretest-posttest trial.
\newblock \emph{The American Statistician}, 55\penalty0 (4):\penalty0 314--321, 2001.

\bibitem[Ye et~al.(2023)Ye, Shao, Yi, and Zhao]{ye2023toward}
Ting Ye, Jun Shao, Yanyao Yi, and Qingyuan Zhao.
\newblock Toward better practice of covariate adjustment in analyzing randomized clinical trials.
\newblock \emph{Journal of the American Statistical Association}, 118\penalty0 (544):\penalty0 2370--2382, 2023.

\bibitem[Williamson et~al.(2014)Williamson, Forbes, and White]{williamson2014variance}
Elizabeth~J Williamson, Andrew Forbes, and Ian~R White.
\newblock Variance reduction in randomised trials by inverse probability weighting using the propensity score.
\newblock \emph{Statistics in medicine}, 33\penalty0 (5):\penalty0 721--737, 2014.

\bibitem[Moore and van~der Laan(2009)]{Moore2009a}
K.L. Moore and M.J. van~der Laan.
\newblock Covariate adjustment in randomized trials with binary outcomes: Targeted maximum likelihood estimation.
\newblock \emph{Stat. Med.}, 28\penalty0 (1):\penalty0 39--64, 2009.
\newblock ISSN 1097-0258.

\bibitem[Chernozhukov et~al.(2018)Chernozhukov, Chetverikov, Demirer, Duflo, Hansen, Newey, and Robins]{chernozhukov2018double}
Victor Chernozhukov, Denis Chetverikov, Mert Demirer, Esther Duflo, Christian Hansen, Whitney Newey, and James Robins.
\newblock Double/debiased machine learning for treatment and structural parameters.
\newblock \emph{The Econometrics Journal}, 21\penalty0 (1):\penalty0 C1--C68, 2018.

\bibitem[Hastie et~al.(2015)Hastie, Tibshirani, and Wainwright]{hastie2015statistical}
Trevor Hastie, Robert Tibshirani, and Martin Wainwright.
\newblock Statistical learning with sparsity.
\newblock \emph{Monographs on statistics and applied probability}, 143\penalty0 (143):\penalty0 8, 2015.

\bibitem[Breiman(2001)]{breiman2001random}
Leo Breiman.
\newblock Random forests.
\newblock \emph{Machine learning}, 45\penalty0 (1):\penalty0 5--32, 2001.

\bibitem[Chipman et~al.(2010)Chipman, George, and McCulloch]{chipman2010bart}
Hugh~A Chipman, Edward~I George, and Robert~E McCulloch.
\newblock Bart: Bayesian additive regression trees.
\newblock \emph{The Annals of Applied Statistics}, pages 266--298, 2010.

\bibitem[Wang et~al.(2019)Wang, Ogburn, and Rosenblum]{wang2019analysis}
Bingkai Wang, Elizabeth~L Ogburn, and Michael Rosenblum.
\newblock Analysis of covariance in randomized trials: More precision and valid confidence intervals, without model assumptions.
\newblock \emph{Biometrics}, 75\penalty0 (4):\penalty0 1391--1400, 2019.

\bibitem[Keele and Small(2021)]{keele2021comparing}
Luke Keele and Dylan~S Small.
\newblock Comparing covariate prioritization via matching to machine learning methods for causal inference using five empirical applications.
\newblock \emph{The American Statistician}, 75\penalty0 (4):\penalty0 355--363, 2021.

\bibitem[Doutreligne and Varoquaux(2023)]{doutreligne2023select}
Matthieu Doutreligne and Ga{\"e}l Varoquaux.
\newblock How to select predictive models for causal inference?
\newblock \emph{arXiv preprint arXiv:2302.00370}, 2023.

\bibitem[Chen et~al.(2025)Chen, Kaufman, Rana, Benmarhnia, and Chen]{chen2025we}
Chen Chen, Jay~S Kaufman, Juwel Rana, Tarik Benmarhnia, and Hong Chen.
\newblock Do we need flexible machine-learning algorithms to assess the effect of long-term exposure to fine particulate matter on mortality?: An example from a canadian national cohort.
\newblock \emph{Environmental Epidemiology}, 9\penalty0 (2):\penalty0 e375, 2025.

\bibitem[Mooney et~al.(2021)Mooney, Keil, and Westreich]{mooney2021thirteen}
Stephen~J Mooney, Alexander~P Keil, and Daniel~J Westreich.
\newblock Thirteen questions about using machine learning in causal research (you won’t believe the answer to number 10!).
\newblock \emph{American Journal of Epidemiology}, 190\penalty0 (8):\penalty0 1476--1482, 2021.

\bibitem[Zivich and Breskin(2021)]{zivich2021machine}
Paul~N Zivich and Alexander Breskin.
\newblock Machine learning for causal inference: on the use of cross-fit estimators.
\newblock \emph{Epidemiology}, 32\penalty0 (3):\penalty0 393--401, 2021.

\bibitem[Hines et~al.(2022)Hines, Dukes, Diaz-Ordaz, and Vansteelandt]{hines2022demystifying}
Oliver Hines, Oliver Dukes, Karla Diaz-Ordaz, and Stijn Vansteelandt.
\newblock Demystifying statistical learning based on efficient influence functions.
\newblock \emph{The American Statistician}, 76\penalty0 (3):\penalty0 292--304, 2022.

\bibitem[Wang et~al.(2024)Wang, Park, Small, and Li]{wang2023CRT}
Bingkai Wang, Chan Park, Dylan~S Small, and Fan Li.
\newblock Model-robust and efficient covariate adjustment for cluster-randomized experiments.
\newblock \emph{Journal of the American Statistical Association}, 119\penalty0 (548):\penalty0 2959--2971, 2024.

\bibitem[Baiardi and Naghi(2024)]{baiardi2024value}
Anna Baiardi and Andrea~A Naghi.
\newblock The value added of machine learning to causal inference: Evidence from revisited studies.
\newblock \emph{The Econometrics Journal}, 27\penalty0 (2):\penalty0 213--234, 2024.

\bibitem[Tsiatis(2006)]{Tsiatis2006}
Anastasios~A. Tsiatis.
\newblock \emph{Semiparametric Theory and Missing Data}.
\newblock Springer, New York, 2006.

\bibitem[Kauermann and Carroll(2001)]{kauermann2001note}
G{\"o}ran Kauermann and Raymond~J Carroll.
\newblock A note on the efficiency of sandwich covariance matrix estimation.
\newblock \emph{Journal of the American Statistical Association}, 96\penalty0 (456):\penalty0 1387--1396, 2001.

\bibitem[Rawshani et~al.(2019)Rawshani, Rawshani, Sattar, Franz{\'e}n, McGuire, Eliasson, Svensson, Zethelius, Miftaraj, Rosengren, et~al.]{rawshani2019relative}
Aidin Rawshani, Araz Rawshani, Naveed Sattar, Stefan Franz{\'e}n, Darren~K McGuire, Bj{\"o}rn Eliasson, Ann-Marie Svensson, Bj{\"o}rn Zethelius, Mervete Miftaraj, Annika Rosengren, et~al.
\newblock Relative prognostic importance and optimal levels of risk factors for mortality and cardiovascular outcomes in type 1 diabetes mellitus.
\newblock \emph{Circulation}, 139\penalty0 (16):\penalty0 1900--1912, 2019.

\bibitem[Li et~al.(2018)Li, Morgan, and Zaslavsky]{li2018balancing}
Fan Li, Kari~Lock Morgan, and Alan~M Zaslavsky.
\newblock Balancing covariates via propensity score weighting.
\newblock \emph{Journal of the American Statistical Association}, 113\penalty0 (521):\penalty0 390--400, 2018.

\bibitem[Hernán and Robins(2020)]{hernan2020causal}
Miguel~A. Hernán and James~M. Robins.
\newblock \emph{Causal Inference: What If}.
\newblock Chapman \& Hall/CRC, Boca Raton, FL, 2020.

\bibitem[Little and Rubin(2019)]{little2019statistical}
Roderick~JA Little and Donald~B Rubin.
\newblock \emph{Statistical analysis with missing data}.
\newblock John Wiley \& Sons, 2019.

\bibitem[Daniels and Hogan(2008)]{daniels2008missing}
Michael~J. Daniels and Joseph~W. Hogan.
\newblock \emph{Missing Data in Longitudinal Studies: Strategies for Bayesian Modeling and Sensitivity Analysis}.
\newblock Chapman \& Hall/CRC, Boca Raton, FL, 2008.

\bibitem[Bugni et~al.(2018)Bugni, Canay, and Shaikh]{bugni2018inference}
Federico~A Bugni, Ivan~A Canay, and Azeem~M Shaikh.
\newblock Inference under covariate-adaptive randomization.
\newblock \emph{Journal of the American Statistical Association}, 113\penalty0 (524):\penalty0 1784--1796, 2018.

\bibitem[Wang et~al.(2023)Wang, Susukida, Mojtabai, Amin-Esmaeili, and Rosenblum]{wang2023model}
Bingkai Wang, Ryoko Susukida, Ramin Mojtabai, Masoumeh Amin-Esmaeili, and Michael Rosenblum.
\newblock Model-robust inference for clinical trials that improve precision by stratified randomization and covariate adjustment.
\newblock \emph{Journal of the American Statistical Association}, 118\penalty0 (542):\penalty0 1152--1163, 2023.

\bibitem[Bannick et~al.(2025)Bannick, Shao, Liu, Du, Yi, and Ye]{bannick2025general}
Marlena~S Bannick, Jun Shao, Jingyi Liu, Yu~Du, Yanyao Yi, and Ting Ye.
\newblock A general form of covariate adjustment in clinical trials under covariate-adaptive randomization.
\newblock \emph{Biometrika}, 112:\penalty0 asaf029, 2025.

\end{thebibliography}
}

\end{document}


\def\spacingset#1{\renewcommand{\baselinestretch}%
{#1}\small\normalsize} \spacingset{1}

\date{\vspace{-5ex}}

\maketitle


\spacingset{1.5}
\setcounter{page}{1}
\appendix

\section{Data curation pipline and trial-level information}
\begin{figure}[htbp]
    \centering
    \includegraphics[width=\linewidth]{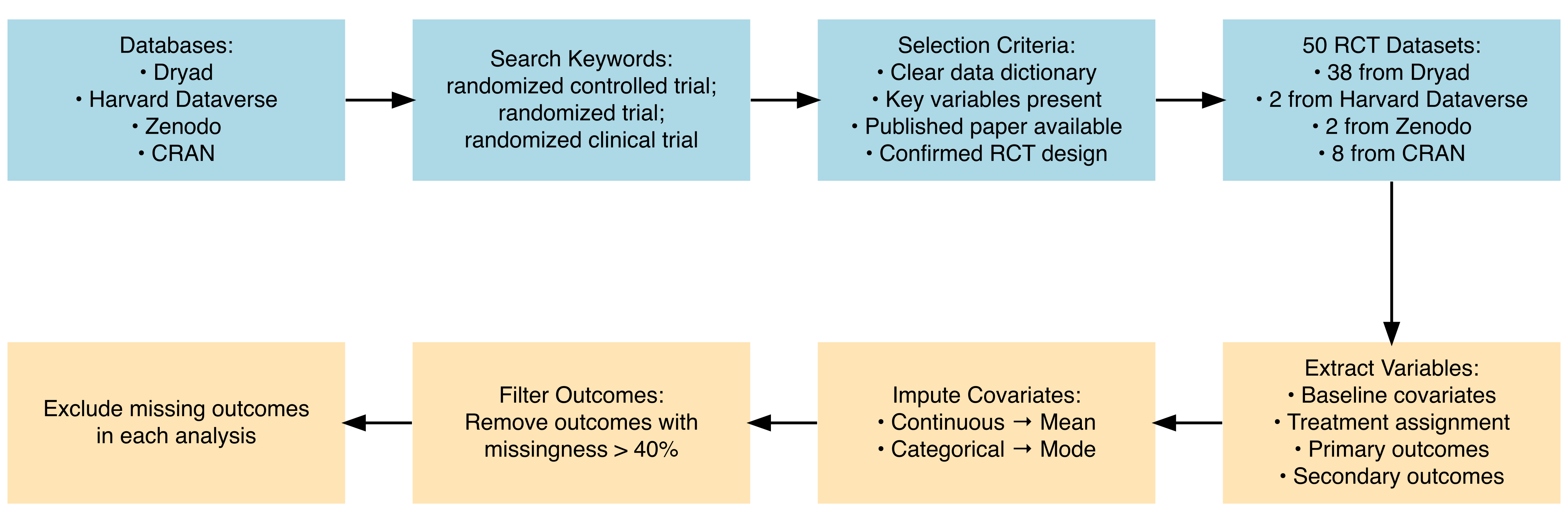}
    \caption{Flowchart of data curation.}
    \label{fig:flowchart}
\end{figure}

Supplement Figure~\ref{fig:flowchart} summarizes the flowchart of data curation, and Supplement Table~\ref{tab:meta_data} lists the main publication name, journal, registration number, number of treatment groups, sample size, research area, and number of covariates for each trial. 


\setlength{\LTleft}{0pt}
\setlength{\LTright}{0pt}
\tiny
\begin{longtable}{p{4.5cm}p{2cm}lllp{2cm}l}
\caption{Trial-level information of 50 RCT datasets. For each dataset, the information includes the title and the journal of the main publication, the registry number (if available), the number of treatment groups, the sample size, the research area, and the number of baseline covariates.} \\
\label{tab:meta_data} \\
\toprule
\textbf{Article Title [Ref.]} & \textbf{Journal} & \textbf{Registry No.} & \textbf{Arms} & \textbf{N} & \textbf{Area} & \textbf{No. Cov.} \\
\midrule
\endfirsthead

\multicolumn{7}{c}{{\bfseries \tablename\ \thetable{} -- continued from previous page}} \\
\toprule
\textbf{Article Title [Ref.]} & \textbf{Journal} & \textbf{Registry No.} & \textbf{Arms} & \textbf{N} & \textbf{Area} & \textbf{No. Cov.} \\
\midrule
\endhead

\midrule \multicolumn{7}{r}{{Continued on next page}} \\
\endfoot
\bottomrule
\endlastfoot

Effects of acupuncture and metformin on insulin sensitivity in women with polycystic ovary syndrome and insulin resistance: a three-armed randomized controlled trial \cite{wen2022effect} & Human Reprod & NCT02491333 & 3 & 342 & Endocrinology & 24 \\

Ivermectin in combination with doxycycline for treating COVID-19 symptoms: a randomized trial \cite{mahmud2021ivermectin} & J Int Med Res & NCT04523831 & 2 & 400 & Infectious Disease & 8 \\

Electronic health record alerts for acute kidney injury: a multicenter, randomized clinical trial \cite{wilson2021electronic} & BMJ & NCT02753751 & 2 & 6030 & Nephrology & 25 \\

Effect of vitamin D supplementation in patients with chronic hepatitis C after direct-acting antiviral treatment: a randomized, double-blind, placebo-controlled trial \cite{sriphoosanaphan2021effect} & PeerJ & TCTR20171206003 & 2 & 75 & Hepatology & 13 \\

A randomized clinical trial assessing the effect of automated medication-targeted alerts on acute kidney injury outcomes \cite{wilson2023randomized} & Nat Commun & NCT02771977 & 2 & 5060 & Nephrology & 42 \\

The effect of sitagliptin on carotid artery atherosclerosis in type 2 diabetes: the PROLOGUE randomized controlled trial \cite{oyama2016effect} & PLoS Med & UMIN000004490 & 2 & 463 & Cardiovascular & 21 \\

Comparing two methods of delivering ThinkRx cognitive training to children ages 8--14: a randomized controlled trial of equivalency \cite{moore2019comparing} & J Cogn Enhanc & NCT02927197 & 2 & 38 & Neurology & 21 \\

Evaluation of ivermectin as a potential treatment for mild to moderate COVID-19: a double-blind, randomized, placebo-controlled trial in Eastern India \cite{ravikirti2021evaluation} & J Pharm Pharm Sci & CTRI/2020/08/027225 & 2 & 115 & Infectious Disease & 3 \\

Umbilical vein oxytocin for the treatment of retained placenta (Release Study): a double-blind, randomised controlled trial \cite{weeks2010umbilical} & Lancet & ISRCTN13204258 & 2 & 577 & Maternal \& Child Health & 8 \\

Influence of videolaryngoscopy using McGrath Mac on the need for a helper to perform intubation during general anaesthesia: a multicentre randomised video-no-video trial \cite{belze2022influence} & BMJ Open & NCT02926144 & 2 & 256 & Anesthesiology & 14 \\

Chlorhexidine oral rinses for symptomatic COPD: a randomised, blind, placebo-controlled preliminary study \cite{pragman2021chlorhexidine} & BMJ Open & NCT02252588 & 2 & 44 & Pulmonology & 14 \\

Effect of an intensive food-as-medicine program on health and health care use \cite{doyle2024effect} & JAMA Intern Med & NCT03718832 & 2 & 349 & Nutrition & 31 \\

Does route matter? Impact of route of oxytocin administration on postpartum bleeding: a double-blind, randomized controlled trial \cite{durocher2019does} & PLoS One & NCT02954068 & 2 & 480 & Maternal \& Child Health & 5 \\

A telerehabilitation programme in post-discharge COVID-19 patients (TERECO): a randomised controlled trial \cite{li2022telerehabilitation} & Thorax & ChiCTR2000031834 & 2 & 119 & Rehabilitation & 24 \\

The effect of Snoezelen intervention on problem behaviors in children with cerebral palsy: a randomized controlled trial \cite{kim2025effect} & Complement Ther Med & KCT0002794 & 2 & 28 & Pediatric Rehab & 17 \\

Internet-delivered therapist-guided physical activity for mild to moderate depression: a randomized controlled trial \cite{strom2013internet} & PeerJ & NCT01573130 & 2 & 48 & Mental Health & 11 \\

The effectiveness and safety of intensive lipid-lowering with different rosuvastatin-based regimens in patients at high cardiovascular disease risk: a nonblind, randomized, controlled trial \cite{lin2023effectiveness} & Rev Cardiovasc Med & ChiCTR2200058389 & 4 & 294 & Cardiovascular & 10 \\

The Kanyakla study: randomized controlled trial of a microclinic social network intervention for promoting engagement and retention in HIV care in rural western Kenya \cite{hickey2021kanyakla} & PLoS One & NCT02474992 & 2 & 295 & HIV/AIDS & 17 \\

Prospective randomized trial comparing hepatic venous outflow and renal function after conventional versus piggyback liver transplantation \cite{brescia2015prospective} & PLoS One & NCT01707810 & 2 & 32 & Transplantation & 12 \\

Effect of ecological momentary assessment, goal-setting and personalized phone-calls on adherence to interval walking training using the InterWalk application among patients with type 2 diabetes: a pilot randomized controlled trial \cite{valentiner2019effect} & PLoS One & NCT02089477 & 2 & 37 & Type 2 Diabetes & 13 \\

Computer-aided X-ray screening for tuberculosis and HIV testing among adults with cough in Malawi (the PROSPECT study): a randomised trial and cost-effectiveness analysis \cite{macpherson2021computer} & PLoS Med & NCT03519425 & 3 & 1462 & HIV/AIDS & 10 \\

Development and first phase evaluation of a maternity leave educational tool for pregnant, working women in California \cite{kurtovich2015development} & PLoS One & -- & 2 & 146 & Maternal \& Child Health & 21 \\

Can recombinant human thrombomodulin increase survival among patients with severe septic-induced disseminated intravascular coagulation: a single-centre, open-label, randomised controlled trial \cite{hagiwara2016can} & BMJ Open & UMIN000008339 & 2 & 92 & Critical Care & 5 \\

Auriculotherapy in the prevention of postoperative urinary retention in patients with thoracotomy and thoracic epidural analgesia: a randomized, double-blinded trial \cite{michel2019auriculotherapy} & Medicine & NCT02290054 & 2 & 50 & Anesthesiology & 12 \\

Multi-strain probiotics (Hexbio) containing MCP BCMC strains improved constipation and gut motility in Parkinson’s disease: a randomised controlled trial \cite{ibrahim2020multi} & PLoS One & NCT04451096 & 2 & 55 & Neurology & 26 \\

Mobile phone reminders and peer counseling improve adherence and treatment outcomes of patients on ART in Malaysia: a randomized clinical trial \cite{abdulrahman2017mobile} & PLoS One & NCT02677675 & 2 & 242 & HIV/AIDS & 9 \\

Impact of the modality of mechanical ventilation on bleeding during pituitary surgery: a single-blinded randomized trial \cite{le2019impact} & Medicine & NCT01891838 & 2 & 101 & Anesthesiology & 4 \\

Effectiveness of community-based health education and home support program to reduce blood pressure among patients with uncontrolled hypertension in Nepal: a cluster-randomized trial \cite{khanal2021effectiveness} & PLoS One & NCT02981251 & 2 & 120 & Cardiovascular & 10 \\

Adjunctive sertraline in HIV-associated cryptococcal meningitis: a randomised, placebo-controlled, double-blind phase 3 trial \cite{rhein2018adjunctive} & Lancet Infect Dis & NCT01802385 & 2 & 460 & HIV/AIDS & 18 \\

A randomized controlled safety and feasibility trial of floatation-REST in anxious and depressed individuals \cite{garland2023randomized} & MedRxiv & NCT03899090 & 3 & 75 & Mental Health & 13 \\

Influenza hemagglutination-inhibition antibody titer as a mediator of vaccine-induced protection for influenza B \cite{cowling2019influenza} & Clin Infect Dis & -- & 2 & 796 & Infectious Disease & 5 \\

Nasal ventilation and rapid maxillary expansion (RME): a randomized trial \cite{iwasaki2021nasal} & Eur J Orthod & ACTRN12617001136392 & 3 & 57 & Orthodontics & 13 \\

Reducing therapeutic misconception: a randomized intervention trial in hypothetical clinical trials \cite{christopher2017reducing} & PLoS One & -- & 2 & 154 & Research Ethics & 8 \\

A randomized synbiotic trial to prevent sepsis among infants in rural India \cite{panigrahi2017randomized} & Nature & NCT01214473 & 2 & 4556 & Maternal \& Child Health & 17 \\

Exercise training and weight gain in obese pregnant women: a randomized controlled trial (ETIP Trial) \cite{garnaes2016exercise} & PLoS Med & NCT01243554 & 2 & 91 & Maternal \& Child Health & 9 \\

Targeting brain health in subjective cognitive decline: insights from a multidomain randomized controlled trial \cite{rolandi2025targeting} & Aging Clin Exp Res & NCT03382353 & 3 & 114 & Neurology & 10 \\

A randomized trial of rectal indomethacin to prevent post-ERCP pancreatitis \cite{elmunzer2012randomized} & N Engl J Med & NCT00820612 & 2 & 602 & Gastroenterology & 28 \\

A trial comparing nucleoside monotherapy with combination therapy in HIV-infected adults with CD4 cell counts from 200 to 500 per cubic millimeter: AIDS Clinical Trials Group Study 175 \cite{hammer1996trial} & N Engl J Med & -- & 4 & 2139 & HIV/AIDS & 16 \\

Linking alcohol- and drug-dependent adults to primary medical care: a randomized controlled trial of a multidisciplinary health intervention in a detoxification unit \cite{samet2003linking} & Addiction & -- & 2 & 470 & Mental Health & 19 \\

A randomized comparison between the Pentax AWS video laryngoscope and the Macintosh laryngoscope in morbidly obese patients \cite{abdallah2011randomized} & Anesth Analg & -- & 2 & 99 & Anesthesiology & 5 \\

A randomized, double-blind comparison of licorice versus sugar-water gargle for prevention of postoperative sore throat and postextubation coughing \cite{ruetzler2013randomized} & Anesth Analg & NCT01444703 & 2 & 235 & Anesthesiology & 8 \\

Combined versus sequential injection of mepivacaine and ropivacaine for supraclavicular nerve blocks \cite{roberman2011combined} & Reg Anesth Pain Med & -- & 2 & 103 & Anesthesiology & 6 \\

Treatment of periodontal disease and the risk of preterm birth \cite{michalowicz2006treatment} & N Engl J Med & NCT00066131 & 2 & 823 & Maternal \& Child Health & 36 \\

Influence of needle-insertion depth on epidural spread and clinical outcomes in caudal epidural injections: a randomized clinical trial \cite{park2018influence} & J Pain Res & NCT03057197 & 2 & 127 & Anesthesiology & 8 \\

Impact of a randomized controlled trial of discounts on fruits, vegetables, and noncaloric beverages in NYC supermarkets on food intake and health risk factors \cite{nzesi2023impact} & PLoS One & NCT04178824 & 3 & 64 & Nutrition & 38 \\

Evaluating the effectiveness of a structured, simulator-assisted, peer-led training on cardiovascular physical examination in third-year medical students: a prospective, randomized, controlled trial \cite{kronschnabl2021evaluating} & GMS J Med Educ & -- & 2 & 70 & Medical Education & 9 \\

Comparison between high-flow nasal cannula (HFNC) therapy and noninvasive ventilation (NIV) in children with acute respiratory failure by bronchiolitis: a randomized controlled trial \cite{santos2024comparison} & BMC Pediatr & U1111-1262-1740 & 2 & 252 & Pediatrics & 17 \\

The effectiveness of breakfast recommendations on weight loss: a randomized controlled trial \cite{dhurandhar2014effectiveness} & Am J Clin Nutr & NCT01781780 & 3 & 255 & Nutrition & 7 \\

Impact of a pre-feeding oral stimulation program on first feed attempt in preterm infants: a double-blind controlled clinical trial \cite{da2020impact} & PLoS One & NCT03025815 & 2 & 74 & Maternal \& Child Health & 26 \\

A controlled trial of interferon gamma to prevent infection in chronic granulomatous disease \cite{international1991controlled} & N Engl J Med & -- & 2 & 128 & Immunology & 8 \\

\end{longtable}
\normalsize

\section{Technical details and implementation of statistical methods}
\subsection{Technical details}
We first introduce several definitions. For individual $i=1,\dots,n$ in an RCT, we define $Y_i$ as the outcome, $A_i$ as the treatment indicator, and $X_i$ as a vector of covariates. We assume that each individual-level data vector $O_i=(Y_i, A_i, X_i)$ is an independent draw from an unknown distribution $\mathcal{P}$. Our goal is to estimate the average treatment effect $\Delta = E[Y|A=1]-E[Y|A=0]$. 

The unadjusted estimator is defined as 
\begin{align*}
    \widehat{\Delta}_{unadj} = \frac{\sum_{i=1}^n A_iY_i}{\sum_{i=1}^n A_i} - \frac{\sum_{i=1}^n (1-A_i)Y_i}{\sum_{i=1}^n (1-A_i)}.
\end{align*}

The ANCOVA estimator is $\widehat{\beta}_A$ obtained by fitting linear regression with ordinary least squares: $E[Y|A,X] = \beta_0 + \beta_A A + \beta_X X$.

The ANCOVA2/ANHECOVA estimator is $\widehat{\beta}_A + \widehat{\beta}_{AX}\sum_{i=1}^n X_i/n$ by by fitting linear regression: $E[Y|A,X] = \beta_0 + \beta_A A + \beta_X X + \beta_{AX} AX$. 

The g-logistic estimator (logistic regression with g-computation) is constructed in two steps. First, we fit logistic regression $\textup{logit}(E[Y|A,X]) = \beta_0 + \beta_A A + \beta_X X$ and denote $\widehat{p}(a,x) = \textup{logit}^{-1}(\widehat\beta_0 + \widehat\beta_A a + \widehat\beta_X x)$. Second, we construct $$\widehat{\Delta}_{g-logistic} = \frac{1}{n} \sum_{i=1}^n \{\widehat{p}(1,X_i) - p(0,X_i)\}.$$

The IPW estimator is defined as
\begin{align*}
    \widehat{\Delta}_{IPW} = \frac{1}{n}\sum_{i=1}^n \left\{\frac{A_iY_i}{\widehat{w}_i} - \frac{(1-A_i)Y_i}{1-\widehat{w}_i}\right\},
\end{align*}
where $\widehat{w}_i = \textup{logit}^{-1}(\widehat{\alpha}_0 + \widehat{\alpha}_X X_i)$ with $(\widehat{\alpha}_0, \widehat{\alpha}_X)$ obtained from fitting the propensity score model $\textup{logit} P(A=1|X) = \alpha_0 + \alpha_X X$.

The unadjusted, ANCOVA, ANCOVA2, g-logistic, and IPW estimators can all be written as Z-estimators (or called M-estimators). Specifically, each estimator can be obtained by solving $\sum_{i=1}^n \psi(O_i;\theta) = 0$, where $\psi$ is the designed estimating function (e.g., score function in maximum likelihood estimation) and $\theta$ is the vector of parameters (e.g., $\beta_0, \beta_A, \beta_X$ in ANCOVA). This expression yields the sandwich variance estimator:
    \begin{align*}
        \left\{\sum_{i=1}^n \frac{\partial}{\partial \theta} \psi(O_i;\theta) \bigg|_{\theta = \widehat{\theta}}\right\}^{-1} \left\{\sum_{i=1}^m  \psi(O_i;\widehat{\theta})\psi(O_i;\widehat{\theta})^\top \right\} \left\{\sum_{i=1}^m \frac{\partial}{\partial \theta} \psi(O_i;\theta) \bigg|_{\theta = \widehat{\theta}}\right\}^{-1\ \top}.
    \end{align*}
This variance estimator is used for uncertainty quantification. See \cite{Tsiatis2006} for complete details of its construction.

The above estimators all have asymptotic model-robustness. Asymptotic model-robustness means that the estimator is consistent and asymptotically normal as the sample size goes to infinity, even if the working model is arbitrarily misspecified. Furthermore, the asymptotic variance can be consistently estimated by the sandwich variance estimator. This property overcomes the restriction of model-based inference and increases the reliability of these estimators.

In addition, ANCOVA guarantees no asymptotic efficiency loss under equal randomization. It means that, if the allocation ratio is 1:1 between treatment and control, then ANCOVA will be at least as precise as the unadjusted estimator. ANHECOVA possesses a similar property: it guarantees no asymptotic efficiency loss given any randomization ratio and becomes asymptotically equivalent to ANCOVA under equal randomization. Finally, IPW has been proven to be asymptotically equivalent to ANHECOVA.

Targeted minimum loss estimation (TMLE) estimates the ATE in randomized trials by combining an initial outcome regression with a targeted update that directly optimizes estimation of the causal parameter. First, TMLE fits a model for the expected outcome given treatment and baseline covariates, often using flexible or machine-learning methods. It then performs a targeted fluctuation step that updates this initial outcome model along a least-favorable submodel so that the resulting estimator solves the efficient influence-function estimating equation for the ATE. The final TMLE is a substitution estimator that averages predicted counterfactual outcomes across treatment arms. 

Debiased (or double) machine learning (DML) estimates the ATE by solving estimating equations such that small errors in the nuisance models have only second-order impact on the target parameter. In RCTs, this procedure is equivalent to solving the efficient influence function for the ATE. DML uses machine-learning methods to estimate the outcome regression, and cross-fitting is used to remove in-sample prediction bias. 

For both TMLE and DML, if the outcome model is consistently estimated in $L_2$ norm, then the resulting estimator will be consistent, asymptotically normal, and efficient. Of note, propensity scores are not estimated in randomized trials because the true treatment assignment probability is known by design. 

\subsection{R implementation}

The unadjusted, ANCOVA, ANHECOVA, and g-logistic estimators were implemented using the \texttt{RobinCar} R package \citep{RobinCar}, which provides robust variance estimators to ensure asymptotic validity. 
The IPW estimator was implemented using the \texttt{PSweight} R package \citep{zhou2020psweight} to fit a logistic regression model for propensity scores. For machine learning-based estimators, TMLE was implemented via the \texttt{tmle} R package \citep{gruber2012tmle}. 
For DML, due to the lack of mature software for randomized trial analysis, we developed custom code following the statistical framework of \cite{chernozhukov2018double}. This tailored implementation produced more stable results compared with general-purpose packages such as \texttt{aipw} \citep{zhong2021aipw}.

Outcome modeling in TMLE and DML employed the \texttt{SuperLearner} R package \citep{van2007super}, using an ensemble of \texttt{SL.glm}, \texttt{SL.glmnet}, \texttt{SL.bartMachine} and \texttt{SL.randomForest}. This ensemble was selected based on the stability and precision performance of each individual machine learning algorithm (See Figure~\ref{fig:indivdiual-sl}). Each learner was used under the default setting without parameter tuning. For numerical stability, the propensity score was fitted with sample means (i.e., the proportion of treatment and control). During implementation, results were flagged if a method produced errors but retained if only warnings were encountered.

\subsection{Performance metrics}
Variance comparison is based on the Proportional Variance Reduction (PVR) relative to the unadjusted analysis, defined as
\begin{equation*}
    \textup{PVR} = 1 - \frac{\widehat{V}_*}{\widehat{V}_{unadj}},
\end{equation*}
where $\widehat{V}_{\text{unadj}}$ denotes the variance estimate from the unadjusted analysis and $\widehat{V}_*$ corresponds to the variance estimate from the covariate-adjusted method. A positive PVR indicates precision gain from covariate adjustment. In the context of clinical trials, PVR quantifies the proportional reduction in variance attributable to covariate adjustment, equivalently, the percentage reduction in required sample size to achieve the same statistical power.

For comparing point estimates, we use the Scaled Difference (S-Diff), defined as
\begin{equation*}
  \textup{S-Diff}   = \frac{\widehat{\Delta}_* - \widehat{\Delta}_{unadj}}{\sqrt{\widehat{V}_{unadj}}},
\end{equation*}
where $\widehat{\Delta}_{\text{unadj}}$ is the unadjusted estimator and $\widehat{\Delta}_*$ is the corresponding adjusted estimator. Because the true treatment effect is unknown in real data analyses, S-Diff provides a standardized measure of how much a covariate-adjusted method tends to increase or decrease the estimated treatment effect relative to the unadjusted analysis.

Covariate adjustment gain for each covariate-adjusted estimator is defined as 
\begin{align*}
    \textup{CAG} = \Pr(p_{*} < 0.05 \mid p_{unadj} \ge 0.05),
\end{align*}
where $p_{*}$ and $p_{unadj}$ are the two-sided p-values of the adjusted and adjusted analysis for the null hypothesis $\textup{ATE} = 0$. Similarly, Covariate adjustment loss is defined as
\begin{align*}
    \textup{CAL} = \Pr(p_{*} \ge 0.05 \mid p_{unadj} < 0.05).
\end{align*}

Error rate is the probability that an R program reports errors without returning results, including failures due to singular covariance matrices or issues arising during cross-fitting. Some algorithms issued warnings related to non-convergence or rank deficiency. Because these warnings were not consistently defined or coded across algorithms, we did not report them separately.

\section{Additional results for primary outcome analyses}
\begin{figure}[htb]

\vspace{-0.1in}

    \centering
\includegraphics[width=1\linewidth]{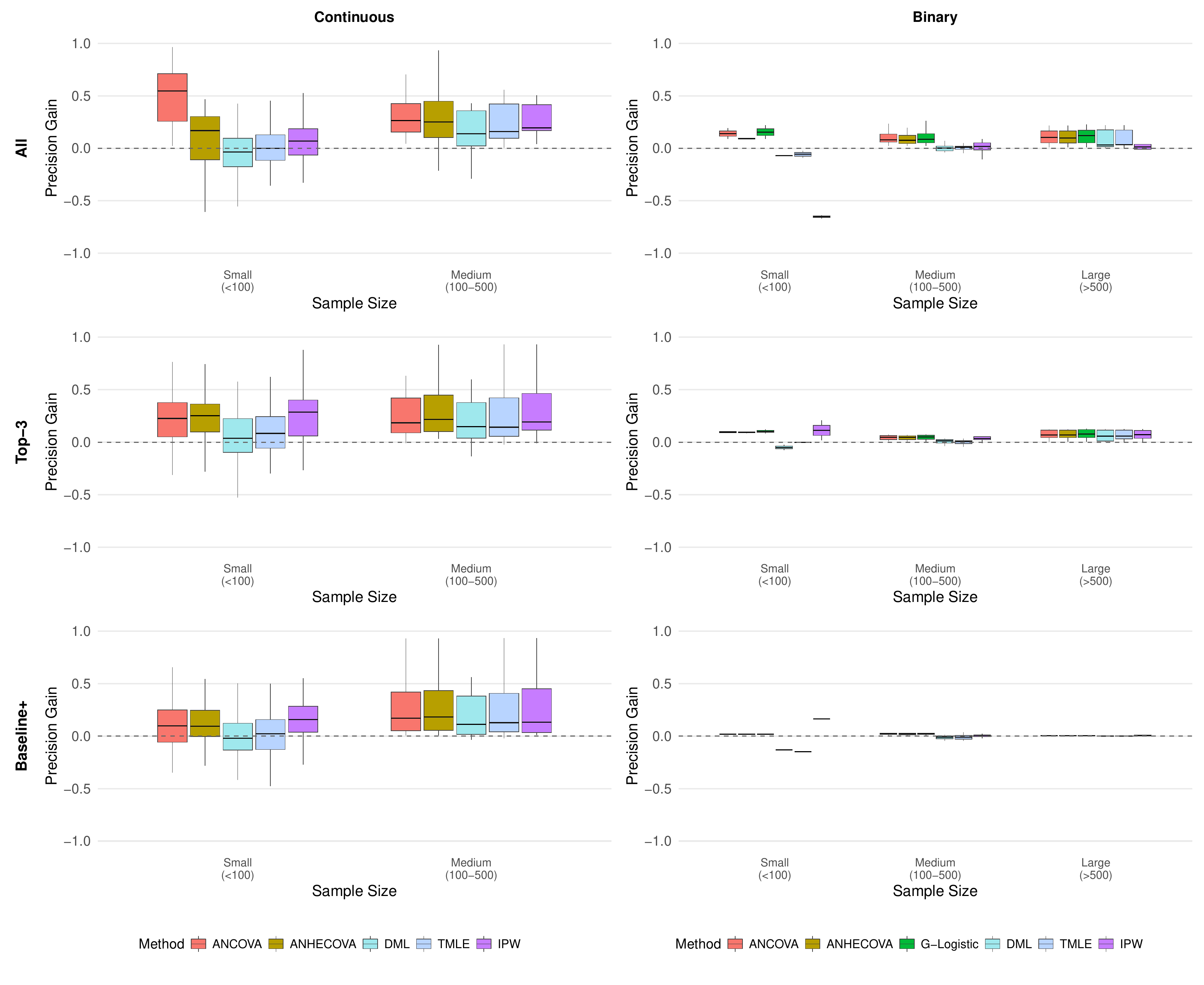}

    \caption{Box plots showing percentage precision gains across different sample-size categories. In the six panels, columns represent outcome types and rows represent covariate-selection strategies, while colors denote the adjustment methods. The y-axis shows the proportional variance reduction (PVR) relative to the unadjusted analysis, with positive values indicating improved precision. The x-axis categorizes trials by sample size: small (0-100), medium (100-500), and large (500+).}
    \label{fig:precision-primary}
\end{figure}

Supplement Figure~\ref{fig:precision-primary} present results for continuous primary outcomes (76 outcome-treatment pairs) and binary primary outcomes (18 outcome-treatment pairs). Overall, the method comparisons exhibit patterns similar to those observed across all 574 outcome-treatment pairs. One notable difference is that covariate adjustment yields smaller precision gains for binary outcomes across settings, which may be attributable to the limited number of binary outcome-treatment pairs.

\section{Comparison of individual machine learning algorithms}
\begin{figure}
    \centering
    \includegraphics[width=\linewidth]{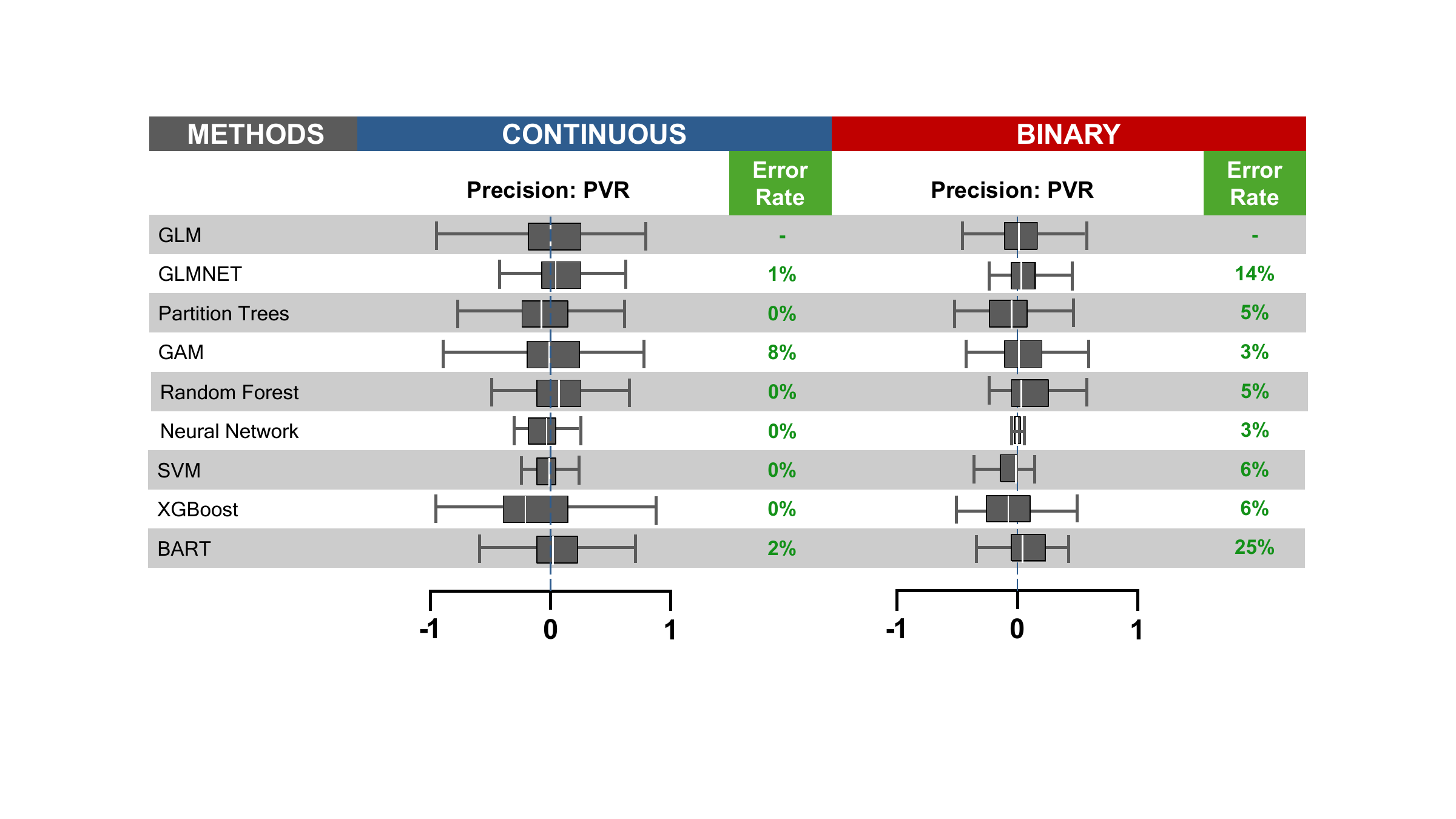}
        \vspace{-1in}

    \caption{Performance of machine-learning algorithms with TMLE adjusting for all covariates. For comparing precision, the box plot of PVR is given, with larger values representing better performance.}
    \label{fig:indivdiual-sl}
\end{figure}

To further assess the performance of various machine learning algorithms, we implemented TMLE with nine algorithms individually and compared their results using PVR and error rate. Figure~\ref{fig:indivdiual-sl} summarizes these findings. In terms of precision gain, only three algorithms, GLMNET, random forest, and BART, yielded positive median PVRs; these are thus included in the ensemble learner. In contrast, the remaining algorithms led to moderate to substantial precision losses, likely due to model misspecification and overfitting. Regarding error rates, most computational failures occurred when cross-fitting with categorical variables that had uncommon levels. In practice, these issues can be mitigated by using an ensemble learner with SuperLearner \citep{van2007super}, which automatically excludes algorithms with fitting errors when multiple learners are used, and by employing stratified cross-validation to ensure all categorical levels are represented.

\section{Additional analysis for the overlap-weighting estimators}
In addition to the estimators presented in the main paper,
we implemented augmented inverse probability weighting (AIPW), overlap weighting (OW), and augmented overlap weighting (AOW) using the \texttt{PSWeight} package \citep{zhou2020psweight}, with generalized linear models for the propensity-score and outcome-regression nuisance functions. 

\begin{figure}[H]
    \centering
    \includegraphics[width=\linewidth]{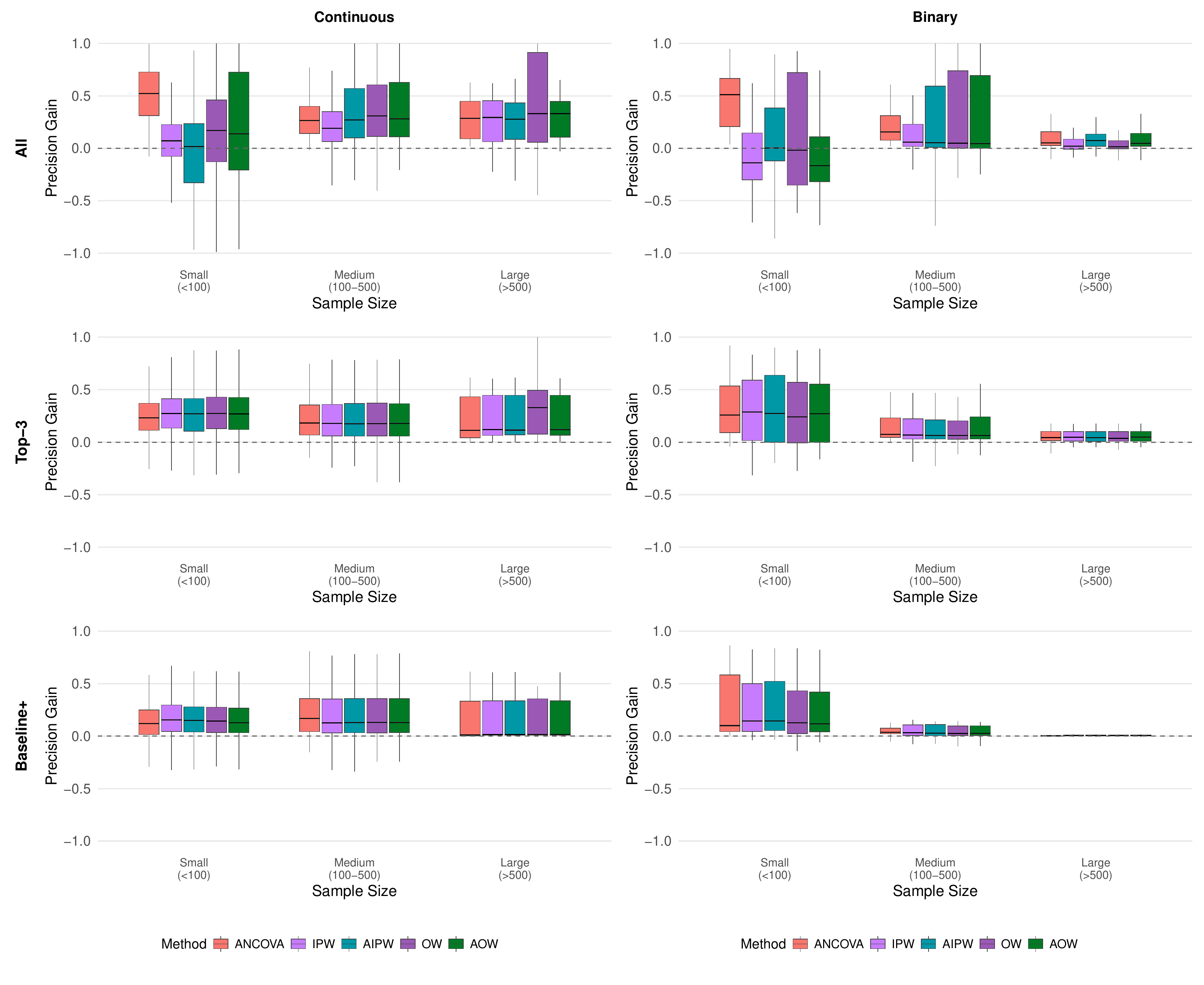}
    \caption{Box plots showing {the empirical distribution of} percentage precision gains across different sample-size categories for the aughmented inverse probability weighting (AIPW), the overlap weighting (OW), and the augmented overlap weighting (AOW) estimators. In the six panels, columns separate outcome types and rows represent covariate-selection strategies, while colors denote the adjustment methods. The y-axis shows the proportional variance reduction (PVR) relative to the unadjusted analysis, with positive values indicating improved precision. The x-axis categorizes trials by sample size: small (0-100), medium (100-500), and large (500+). }
    \label{fig:r1-1}
\end{figure}

\begin{figure}[H]
    \centering
    \includegraphics[width=\linewidth]{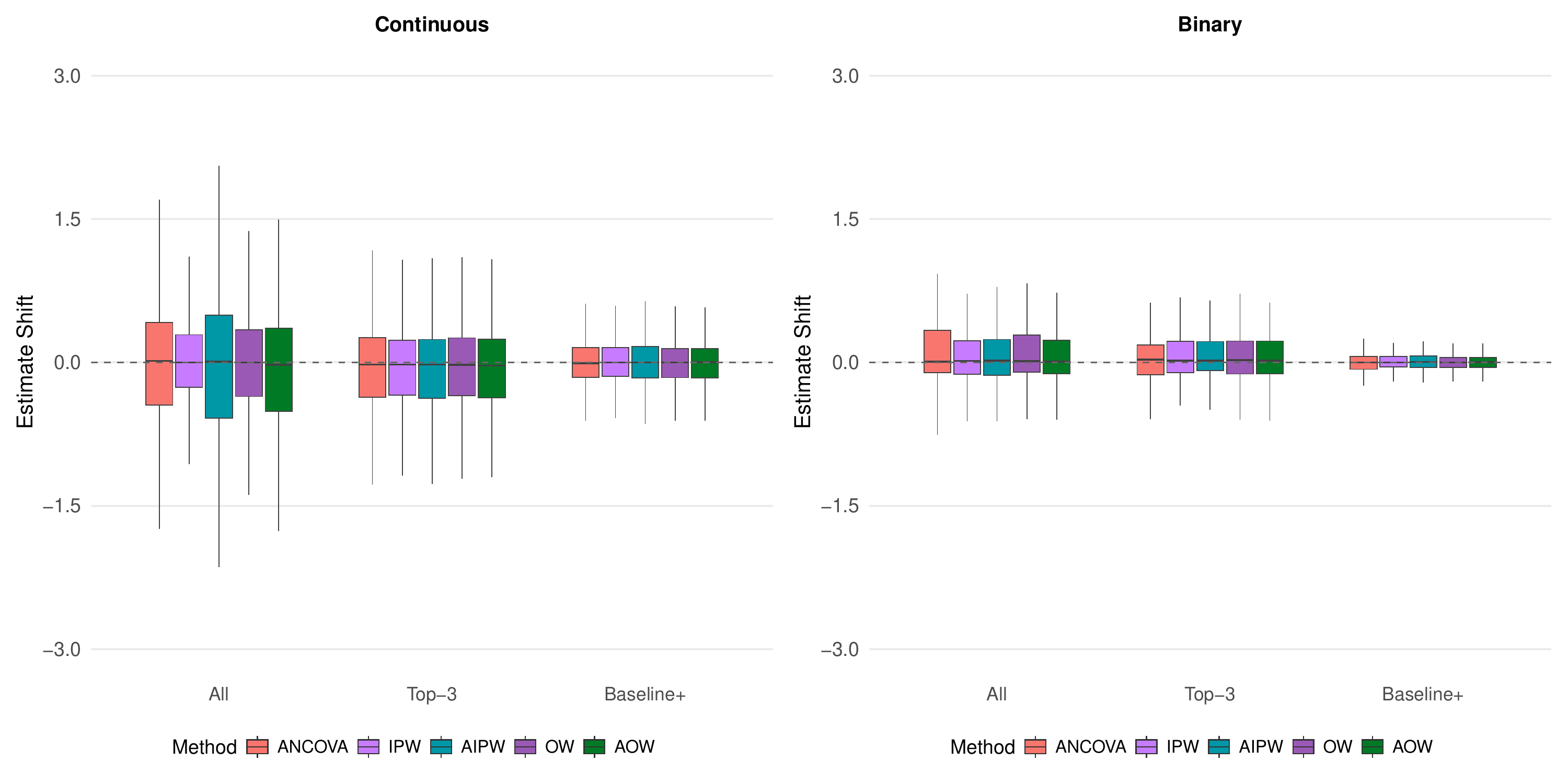}
    \caption{Box plots showing {the empirical distribution of} estimate shifts scaled by the variance of the unadjusted estimators across adjustment methods and covariate-selection strategies for the aughmented inverse probability weighting (AIPW), the overlap weighting (OW), and the augmented overlap weighting (AOW) estimators. {Column panels separate outcome types.}}
    \label{fig:r1-2}
\end{figure}

Figures~\ref{fig:r1-1} and \ref{fig:r1-2} summarized the results, together presented with ANCOVA and IPW for comparison. Overall, overlap weighting outperformed IPW in several settings in terms of precision while maintaining similar estimate shifts. The augmented methods performed similarly to ANCOVA and IPW under the Top-3 and Baseline+ strategies, but showed precision losses when adjusting for all covariates in small samples. We conjecture that this may be partly because the \texttt{PSWeight} package uses conservative variance estimators when the sample size is close to the number of covariates, a situation that arises in several small-sample settings.


\vspace{0.4cm}

{
\bibliography{references}
}